# Energy Efficient Power Allocation in Massive MIMO NOMA Systems Based on SIF Using Cell Division Technique

Abdolrasoul Sakhaei Gharagezlou
*Faculty of Electrical and Computer Engineering*
*University of Tabriz*
Tabri, Iran
Abdolrasoulsakhaei@gmail.com

Jafar Pourrostam
*Faculty of Electrical and Computer Engineering*
*University of Tabriz*
Tabriz, Iran
j.pourrostam@tabrizu.ac.ir

Mahdi Nangir
*Faculty of Electrical and Computer Engineering*
*University of Tabriz*
Tabriz, Iran
nangir@tabrizu.ac.ir

Mir Mahdi Safari
*Faculty of Electrical and Computer Engineering*
*University of Tabriz*
Tabriz,Iran
safari@tabrizu.ac.ir

*Abstract*— In this paper, we investigate energy-efficient power allocation for the downlink of the massive multiple-input multiple-output (MIMO) non-orthogonal multiple access (NOMA) systems. In our proposed scheme, we divide a cell into two zones. The first area is for users whose distance from the base station (BS) is less than half of the radius of cell and the second area is for users whose distance from BS is more than half of the radius of cell. Based on distance of users from BS and the number of users in each area, we dedicate an amount of power for each user. We also use standard interference function (SIF) to propose a new iterative algorithm to solve the optimization problem and obtain the optimal power allocation scheme. Simulation results show that the proposed algorithm outperforms other algorithms from the energy efficiency (EE) point of view.

## I. Introduction

Massive multiple-input multiple-output (MIMO) systems have an outstanding performance in terms of the energy efficiency (EE) and the spectral efficiency (SE). For this reason, they are being considered as one of the key technologies in fifth generation (5G) systems. Important features of the MIMO systems include high spectral and energy efficiency, reliable communication and low-complexity signal processing [1]. The effects of small-scale noise and fading are eliminated when the number of antennas in massive MIMO systems goes to infinity. The massive MIMO systems use antenna arrays with more than 100 elements for the direct energy transmission [2]. One of the technologies considered in the 5G networks in combination with massive MIMO systems is the non-orthogonal multiple access (NOMA) method. This combination can provide a significant improvement in the coverage probability and the sum rate in a network [3]. Providing significantly higher spectral efficiency than the Orthogonal Multiple Access (OMA) for several users in a same time, frequency and code, with different power levels is a key feature of the NOMA [4]. There are many works about the EE of massive MIMO NOMA [5] and the references therein. The energy efficient power allocation for a MIMO NOMA system with multiple users in a cluster is investigated to ensure a minimum rate for each user [6]. In [7], a two-step iterative EE algorithm with joint antennas and user selection is proposed in single-cell MIMO systems. The simulation results of this algorithm show that the EE can be improved by 71.16% with the maximum-ratio combining (MRC) receiver when the total number of users is 60. Previous work has focused on the sub-channel selection and power allocation to maximize the sum rate, however, the allocation of efficient energy resources problem for the NOMA systems has not been well studied. The optimization and allocating of the sub-channel power to maximize EE for the NOMA networks in the downlink are discussed in [8]. Most of the relevant works have focused on the effect of the wireless channels. In fact, the performance of the massive MIMO systems in addition to the wireless channel depends on the transceiver Radio Frequency (RF). Therefore, the effect of transceiver RF on the performance of the massive MIMO systems should be considered. In [9], a detailed analysis of the sum rate utilizing maximum ratio transmit (MRT) and Zero Forcing (ZF) pre-coding is presented. A motivation to investigate the efficient power allocation scheme for the massive MIMO systems with the NOMA method is established. Recently, in [10], the energy efficient power allocation for the massive MIMO system with multiple users in one cell is considered. Furthermore, an iterative algorithm is proposed for the optimal power allocation scheme using SIF [10]. In this paper, we propose an energy efficient power allocation scheme in the massive MIMO NOMA scenario motivated by [10]. The properties of our proposed scheme are summarized as follows:

- We divide a cell into two zones. The first zone belongs to users whose distance from the BS is less than half of the radius of the cell .We allocate less amount of the total transmission power to this zone. The second zone belongs to users whose distance from the BS is more than half the radius of the cell, which we allocate a greater amount of the total transmission power to this area. This idea leads to break the original linear power constraint in convex optimization problem in [10], into two separate linear power constraints.

- We propose a new standard interference function (SIF) based iterative algorithm for the cell division senario. The proposed algorithm performs better than other similar power allocation algorithms.

The organization of this paper is as follows. In section II, the system model is described. In section III, the optimization problem is formulated and we obtain the optimal power allocation scheme. In section IV, the simulation results are presented. Finally, the conclusion appears in section V.

## II. SYSTEM MODEL

In this paper, we consider the system depicted in Fig. 1. In this system, the base station (BS) is equipped with $M$ antennas and each user is equipped with single antenna. Consider $K$ number of users share the same Resource Block (RB). $\mathbf{G}$ is the matrix of the flat fading channel between the BS and the $K$ users. $\mathbf{G}$ can be formulated as:

$$\mathbf{G} = \mathbf{H}\mathbf{Q}^{1/2}, \quad (1)$$

where $\mathbf{H} \in C^{M \times K}$ is the small scale fading channel matrix and $\mathbf{Q} = diag\{\beta_1, \beta_2, \ldots, \beta_K\}$ denotes the large scale fading matrix, where $\beta_K = \varphi\theta/d_k^\varepsilon$ represents the path loss and shadow fading. $\varphi$ is a constant related to the carrier frequency and antenna gain, $d_k$ is the distance between the BS and the $k$-th user, $\varepsilon \in [2, 6]$ represents the path loss exponent, and $\theta$ is the shadow fading with lognormal distribution, i.e., $10\log_{10}(\theta) \sim N(0, \sigma^2)$. Accordingly, at user $k$, the observed signal can be formulated as [10]:

$$y_k = \sqrt{p_k}\mathbf{g}_k^H \mathbf{v}_k s_k + \sum_{\mathcal{K}=1, \mathcal{K} \neq k}^{K} \sqrt{p_\mathcal{K}}\mathbf{g}_k^H \mathbf{v}_\mathcal{K} s_\mathcal{K} + n_k, \quad (2)$$

where $k \in \{1, 2, \ldots, K\}$, $p_k$ denotes the transmit power allocated to the $k$-th user, $\mathbf{g}_k$ is the $k$-th column of $\mathbf{G}$, $v_k$ is the precoding column for the $k$-th user, $s_k$ represents transmit data symbol of the $k$-th user, $n_k$ is the additive white Gaussian noise (AWGN) at the $k$-th user with distribution $N(0, N_0)$, and $N_0$ denotes the noise power spectral density.

To balance performance and complexity, we use MRT precoding.

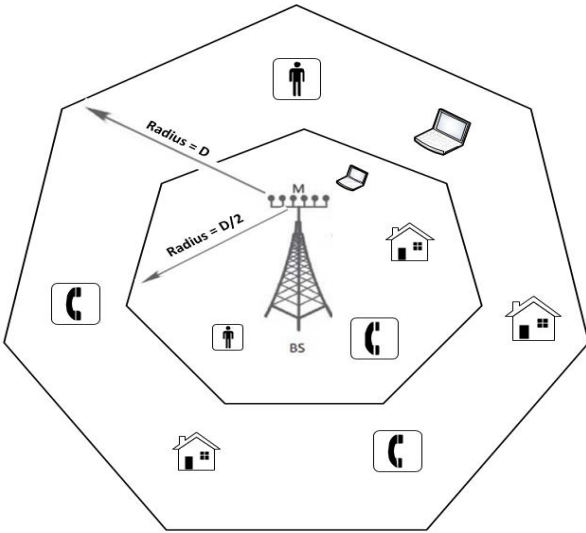

Fig. 1. System model of a multiuser massive MIMO system via Cell Division.

Hence, the precoding vector for the $k$-th user can be formulated as [11]:

$$\mathbf{v}_k = \frac{\mathbf{g}_k}{\|\mathbf{g}_k\|}, \quad (3)$$

where $\|.\|$ represents the $L2$-norm. Let $B$ represents the bandwidth of one RB.

The Signal to Interference and Noise Ratio (SINR) of the $k$-th user for a given $\mathbf{g}_k$ can be formulated as:

$$\gamma_k = \frac{p_k \left|\mathbf{g}_k^H \mathbf{v}_k\right|^2}{\sum_{\mathcal{K}=1, \mathcal{K} \neq k}^{K} p_\mathcal{K} \left|\mathbf{g}_k^H \mathbf{v}_\mathcal{K}\right|^2 + BN_0}. \quad (4)$$

As a result, the achieved data rate at user $k$ is given by:

$$r_k = B\log_2(1 + \gamma_k). \quad (5)$$

We define the EE of the system as [10]:

$$EE = \frac{\sum_{k=1}^{K} r_k}{\sum_{k=1}^{K} p_k + \sum_{m=1}^{M} p_{c,m}}, \quad (6)$$

where $p_{c,m}$ is the fixed circuit power consumption per antenna.

## III. THE PROPSED POWER ALLOCATION ALGORITHM

In this section, we first formulate the proposed energy efficient power allocation scheme and then the method of solving the problem will be explained.

### A. Optimization problem formulation

We aim to maximize the EE of the system when each user has a pre-defined minimum rate. The considered problem can be formulated as:

$$max_{\{p_1, p_2, \ldots, p_k\}} EE \quad (7.a)$$

$$s.t. \quad C_1 : r_k \geq R_{T.k}; \quad k = 1, 2, \ldots, K \quad (7.b)$$

$$C_2 : \sum_{k: d_k < \frac{D}{2}} p_k = \alpha P_T \quad (7.c)$$

$$C_3 : \sum_{k: d_k > \frac{D}{2}} p_k = (1-\alpha)P_T, \quad (7.d)$$

where constraints $C_1$, $C_2$ and $C_3$ represent the users' minimum rate requirements, the transmit power constraint of first zone and the transmit power constraint of second zone respectively. $P_T$ denotes the flexible transmit power and $R_{T.k}$ denotes the minimum data rate requirement for the $k$-th user. $\alpha$ denotes the power allocation coefficient for the first zone, and $D$ is the radius of the cell. It is worth to note that constraint (7.c) means that for users in the first zone, they are less than half the radius of the cell, $\alpha$ times the total power allocated. The optimization problem in (7.a) is a constrained non-convex optimization problem which is formulated to maximize the EE, considering the quality of service requirements.

## B. Proposed solution

The objective function in (6) is in a fractional form. Thus, it can be classified as a non-linear fractional programming. By exploiting the properties of fractional programming and the lower bound of the user data rate, the non-convex optimization problem in (7.a) is transformed into a convex optimization problem. By following a similar approach as in [10], the maximum EE can be achieved if and only if:

$$max_{\{p_1,p_2,...,p_K\}} \left\{ \sum_{k=1}^{K} p_k - q^* \left( \sum_{k=1}^{K} p_k + \sum_{m=1}^{M} P_{c.m} \right) \right\} = 0, \quad (8)$$

where $q^*$ represents maximum EE. By assuming perfect channel state information (CSI), and Rayleigh fading and MRT precoding, the downlink data rate for the $k$-th user can be lower bounded by exploiting the properties of high dimensional channel vector as follows [11]:

$$r_k = B\log_2 \left( 1 + \frac{M \beta_k p_k}{\beta_k \sum_{\mathcal{K}=1, \mathcal{K} \neq k}^{K} p_{\mathcal{K}} + BN_0} \right). \quad (9)$$

In high SINR region, the received SINR in (9) is much larger than 1, therefore, the lower bound of $r_k$ can be formulated as:

$$\tilde{r}_k = B\log_2 \left( \frac{M \beta_k p_k}{\beta_k \sum_{\mathcal{K}=1, \mathcal{K} \neq k}^{K} p_{\mathcal{K}} + BN_0} \right). \quad (10)$$

By using (10), the original optimization problem in (7) can be simplified as follows:

$$max_{\{p_1,p_2,...,p_K\}} \left\{ \sum_{k=1}^{K} \tilde{r}_k - q^* \left( \sum_{k=1}^{K} p_k + \sum_{m=1}^{M} P_{c.m} \right) \right\} \quad (11.a)$$

$$s.t. \quad C_1 : \tilde{r}_k \geq R_{T.k} ; \quad k = 1,2,...,K \quad (11.b)$$

$$C_2 : \sum_{k:d_k < \frac{D}{2}} p_k = \alpha P_T \quad (11.c)$$

$$C_3 : \sum_{k:d_k > \frac{D}{2}} p_k = (1-\alpha) P_T . \quad (11.d)$$

The simplified optimization problem in (11) is a convex optimization problem. From the subtractive EE function in (11.a), it can be seen that the term $q^* \left( \sum_{k=1}^{K} p_k + \sum_{m=1}^{M} P_{c.m} \right)$ is an affine function. On the other hand the sum rate function $\sum_{k=1}^{K} \tilde{r}_k$ is concave [10]. Consequently, the subtractive EE function in (11.a) is concave. Besides, it can be shown that $C_1$ is a convex constraint and $C_2$ and $C_3$ are liner affine constraints. Therefore, we can conclude that the simplified optimization problem in (11) is a convex problem.

The Lagrangian dual function method can be used to convert constrained convex problem to unconstrained convex problem. Let $\Phi$ be the Lagrangian function of (11), then it can be formulated as follows [12]:

$$\Phi(\mathbf{p}, \omega_1, \omega_2, \mathbf{\rho}) = -\left[ \sum_{k=1}^{K} \tilde{r}_k - q \left( \sum_{k=1}^{K} p_k + \sum_{m=1}^{M} P_{c.m} \right) \right]$$
$$- \omega_1 \left( \alpha P_T - \sum_{k:d_k < \frac{D}{2}} p_k \right)$$
$$- \omega_2 \left( (1-\alpha) P_T - \sum_{k:d_k > \frac{D}{2}} p_k \right) \quad (12)$$
$$- \sum_{k=1}^{K} \rho_k \left( \tilde{r}_k - R_{T.k} \right),$$

$\mathbf{p}$ represents feasible set of power variables, $\omega_1, \omega_2 \geq 0$ is the Lagrangian multiplier corresponding to the transmit power constraint of the first zone and the second zone, respectively. $\mathbf{\rho}$ is the Lagrangian multiplier vector corresponding to the data rate constraints with its element $\rho_k \geq 0$. The necessary and sufficient condition to obtain the optimal transmit power for the first zone $k : d_k < D/2$ can be formulated as:

$$\frac{\partial \phi}{\partial p_k} = \sum_{\mathcal{K}=1, \mathcal{K} \neq k} \frac{1+\rho_{\mathcal{K}}}{\left( \sum_{\mathcal{K}'=1, \mathcal{K}' \neq k}^{K} p_{\mathcal{K}'} + \frac{BN_0}{\beta_{\mathcal{K}'}} \right) \ln 2} - \frac{1+\rho_{\mathcal{K}}}{p_{\mathcal{K}} \ln 2} + q + \omega_1 = 0. \quad (13)$$

Also, the optimal transmit power for the second zone $k : d_k > D/2$ can be formulated as:

$$\frac{\partial \phi}{\partial p_k} = \sum_{\mathcal{K}=1, \mathcal{K} \neq k} \frac{1+\rho_{\mathcal{K}}}{\left( \sum_{\mathcal{K}'=1, \mathcal{K}' \neq k}^{K} p_{\mathcal{K}'} + \frac{BN_0}{\beta_{\mathcal{K}'}} \right) \ln 2} - \frac{1+\rho_{\mathcal{K}}}{p_{\mathcal{K}} \ln 2} + q + \omega_2 = 0. \quad (14)$$

From (13) and (14), the optimal transmit power for the $k$-th user can be formulated as follows:

First zone $k : d_k < D/2$:

$$p_k = \frac{1+\rho_k}{\left( \sum_{\mathcal{K}=1, \mathcal{K} \neq k} \frac{1+\rho_{\mathcal{K}}}{\Pi} + q + \omega_1 \right) \ln 2},$$
$$\Pi = \left( \sum_{\mathcal{K}=1, \mathcal{K} \neq k} p_{\mathcal{K}'} + \frac{BN_0}{\beta_{\mathcal{K}'}} \right) \ln 2. \quad (15)$$

Second zone $k : d_k > D/2$:

$$p_k = \frac{1+\rho_k}{\left( \sum_{\mathcal{K}=1, \mathcal{K} \neq k} \frac{1+\rho_{\mathcal{K}}}{\chi} + q + \omega_2 \right) \ln 2},$$
$$\chi = \left( \sum_{\mathcal{K}=1, \mathcal{K} \neq k} p_{\mathcal{K}'} + \frac{BN_0}{\beta_{\mathcal{K}'}} \right) \ln 2. \quad (16)$$

The parameters that are most important in allocating full transmission power to two zones are the number of users in that area and the distance to the base station [13]. The

amount of total transmission power $\alpha$ that is allocated to the first area equals:

$$\alpha = \frac{\sum_{k:d_k<\frac{D}{2}} d_k^2}{\sum_{k=1}^{K} d_k^2}. \quad (17)$$

Therefore, the amount of total transmission power that is allocated to the second area as follows:

$$(1-\alpha) = 1 - \frac{\sum_{k:d_k<\frac{D}{2}} d_k^2}{\sum_{k=1}^{K} d_k^2} = \frac{\sum_{k:d_k>\frac{D}{2}} d_k^2}{\sum_{k=1}^{K} d_k^2}. \quad (18)$$

*C. Proposed SIF-based iterative algorithm*

In this section we propose a new SIF based iterative algorithm. Our proposed algorithm is summarized in **Algorithm 1**. Note, the parameter $\tau$ defines the predetermined positive threshold for the terminating condition, $n$ shows the iteration number, and $\theta_1$ and $\theta_{2,k}$ denote the positive step sizes. The condition $p_k^{(n)} = p_k^{(n+1)}$ guarantees fast convergence of the proposed algorithm. Due to the distance of users from the BS, the power allocation relationship is considered for the two areas. Because the power allocation of users follows different relationships, the Lagrangian coefficient update for total transmission power is also different for these two areas. For more accurate EE, the value of $\tau$ is considered $10^{-5}$.

IV. SIMULATION RESULTS

In this section, we show the performance of the proposed algorithm by interpreting simulation results. A single cell with a radius of 500 meters is considered. Let $P_{c,m} = P_c$ for $m = 1,2,\ldots,M$ and $R_{T,k} = R_T$ for $k = 1,2,\ldots,K$. The parameters used in our simulations are presented in Table 1. The values in the simulation are used in a way that can be comparable to previous works

In Fig. 2, we compare the performance of the proposed algorithm with the reported in [10]. As it can be seen in Fig. 2, the proposed SIF function via cell division algorithm performs better than the SIF function.

TABLE 1. SIMULATION PARAMETERS

| Parameter | Value |
|---|---|
| The radius of the cell $D$ | 500 m |
| RB bandwidth $B$ | 120 kHz |
| Number of transmit antennas $M$ | 128 |
| Number of users $K$ | 3 |
| Variance of log-normal shadow fading $\sigma^2$ | 10 dB |
| Factor $\varphi$ | 1 |
| Noise spectral density $N_0$ | -170 dBm/Hz |
| User rate constraint $R_T$ | 3 bit/s/Hz |

**Algorithm 1:** Proposed SIF-based iterative algorithm.

1: **Initialized the transmit power and Lagrangian multipliers:** $\mathbf{p}^{(0)}, \omega_1^{(0)}, \omega_2^{(0)}, \boldsymbol{\rho}^{(0)}$

2: **Calculate the initial EE** $q^{(0)} = \frac{\sum_{k=1}^{K} \tilde{r}_k^{(0)}}{\sum_{k=1}^{K} p_k^{(0)} + \sum_{m=1}^{M} p_{c,m}}$

3: **while** $\sum_{k=1}^{K} \tilde{r}_k^{(n)} - q^* \left( \sum_{k=1}^{K} p_k + \sum_{m=1}^{M} P_{c,m} \right) > \tau$ **do**

4: **for** $i=1:K$ **do**

If $d_k < \frac{D}{2}$

$$p_k^{(n+1)} = \frac{1+\rho_k}{\left( \sum_{\mathcal{K}=1,\mathcal{K}\neq k} \frac{1+\rho_{\mathcal{K}}}{\left( \sum_{\mathcal{K}'=1,\mathcal{K}'\neq k}^{K} p_{\mathcal{K}'}^{(n)} + \frac{BN_0}{\beta_{\mathcal{K}'}} \right) ln2} + q + \omega_1 \right) ln2}$$

Else if $d_k > \frac{D}{2}$

$$p_k^{(n+1)} = \frac{1+\rho_k}{\left( \sum_{\mathcal{K}=1,\mathcal{K}\neq k} \frac{1+\rho_{\mathcal{K}}}{\left( \sum_{\mathcal{K}'=1,\mathcal{K}'\neq k}^{K} p_{\mathcal{K}'}^{(n)} + \frac{BN_0}{\beta_{\mathcal{K}'}} \right) ln2} + q + \omega_2 \right) ln2}$$

End if
5: **End for**
**Update**
$p_k^{(n)} = p_k^{(n+1)}$

$$\omega_1^{(n+1)} = max\left(0, \omega_1^{(n)} - \theta_1 \left( \alpha P_T - \sum_{k:d_k<\frac{D}{2}} p_k \right)\right)$$

$$\omega_2^{(n+1)} = max\left(0, \omega_2^{(n)} - \theta_1 \left( (1-\alpha) P_T - \sum_{k:d_k>\frac{D}{2}} p_k \right)\right)$$

$$\rho_k^{(n+1)} = max\left(0, \rho_k^{(n)} - \theta_{2,k} \left( \tilde{r}_k^{(n)} - R_{T,k} \right)\right)$$

$$q^{(n+1)} = \frac{\sum_{k=1}^{K} \tilde{r}_k^{(n)}}{\sum_{k=1}^{K} p_k^{(n)} + \sum_{m=1}^{M} p_{c,m}}$$

$n = n+1$.
6: **end while**
7: **End**

In this simulation, it is assumed that the total transmission power is 1 watt .For instance, the energy efficiency of the proposed method is 8.1 Mbit/j for a fixed power value of 2dBm, which has been improved about 0.95 Mbit/j compared to the reported in [10].

In Fig. 3, we present how the number of antennas affects the EE. In this simulation, it is assumed that the constant power of the circuit is 7dBm .Obviously, in Fig. 3, the proposed method outperforms the method which has been reported in [10]. For instance, the energy efficiency of the proposed method is 4.35 Mbit/j for the case of 64 antennas, which has been improved about 0.39 Mbit/j compared to the reported in [10].

Fig .4. shows that transmission power of users in the proposed method is lower than the reported in [10], which increases EE. Furthermore, in Fig. 4, the proposed SIF function via cell division algorithm performs better than the SIF function. In this simulation, it is assumed that the maximum transmission power is 1 watt. The power consumption in the proposed method is 19.75 dBm for the case of 128 antennas, which has been improved about 0.65 dBm compared reported in [10].

In Table 2, we compare the EE versus the maximum transmission power. According to the results in Table 2, the proposed SIF function via cell division algorithm performs better than the SIF function.

In Table 3, we compare the EE versus the user rate constraint. In this simulation, the total transmission power is 1 watt, the constant power of the circuit is 10dBm, and the number of transmissions is 128. This table also shows the amount of improvement. According to the simulation results, we show that the proposed algorithm outperforms the reported in [10] from EE point of view. According to the results of [10], our proposed algorithm is also better than the algorithms which are presented in [11] and [12]. These two works are based on the convex optimization theory and the game theory, respectively.

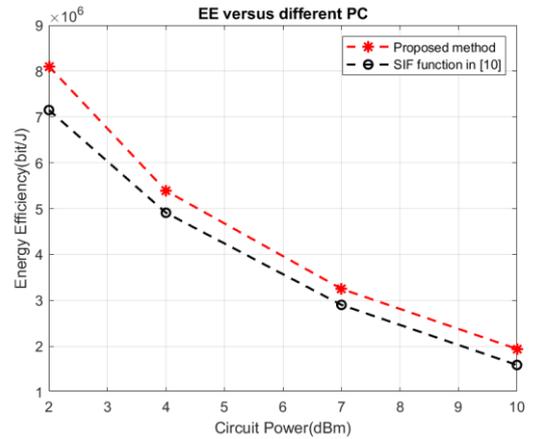

Fig. 2. EE versus different $P_c$

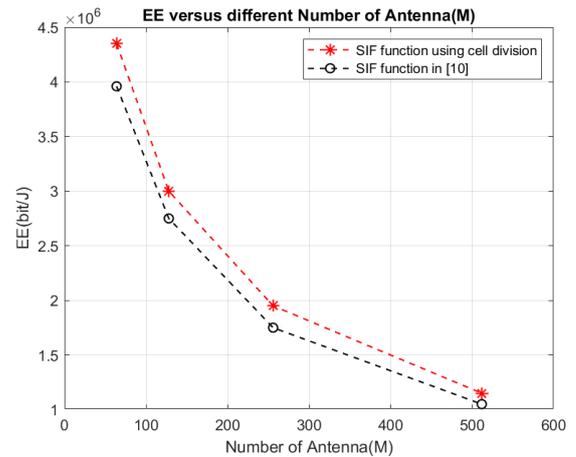

Fig. 3. EE versus different $M$.

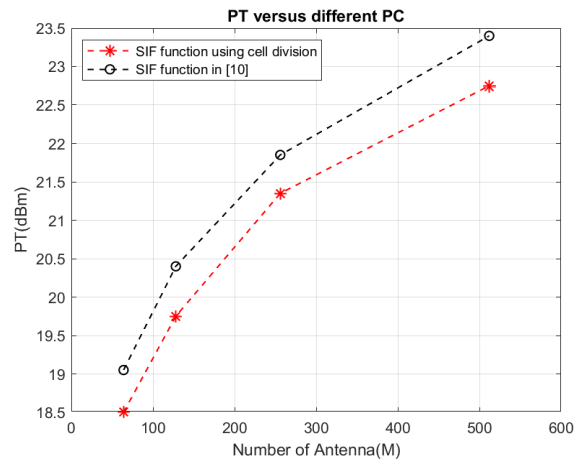

Fig. 4. $P_T$ versus different $M$.

TABLE 2. EE VERSUS THE DIFFERENT $P_T$.

| maximum transmission power (w) | EE convergence propose algorithm (Mbit/J) | EE convergence algorithm in [10] (Mbit/J) | Amount of improvement (Mbit/J) |
|---|---|---|---|
| 2 | 5.4026 | 5.0338 | 0.3688 |
| 3 | 5.4022 | 5.0321 | 0.3701 |
| 4 | 5.4016 | 5.0295 | 0.3720 |

TABLE 3. EE VERSUS THE DIFFERENT $R_T$.

| User rate constraint $R_T$ (bit/s/Hz) | EE convergence propose algorithm (Mbit/J) | EE convergence algorithm in [10] (Mbit/J) | Amount of improvement (Mbit/J) |
|---|---|---|---|
| 4 | 1.6357 | 1.2321 | 0.4036 |
| 5 | 1.5271 | 1.2148 | 0.3123 |
| 6 | 1.4384 | 1.1985 | 0.2399 |

## V. CONCLUSION

In this paper, we formulated the energy efficient power allocation problem for the massive MIMO NOMA system. In the proposed scheme, a major part of the total power is allocated to the users which are farther from the BS. In our scheme, the sum of user transmit power and minimum data rate constraints are taken into consideration. An iterative algorithm is proposed for a two-zone cell based on a multi-criteria standard interference function. According to the presented results, we show that the proposed algorithm performs better than other similar power allocation algorithms.